\documentclass[prl,aps,twocolumn,superscriptaddress,nofootinbib,longbibliography]{revtex4-2}
\usepackage{graphicx} 
\usepackage{dcolumn}  
\usepackage{bm}       
\usepackage{amsmath}
\usepackage{epsfig}
\usepackage{color}

\def\fl{^{{\rm fl}}}

\begin{document}
\title{Near-Field Radiative Heat Transfer Eigenmodes}
\author{Stephen Sanders}
\affiliation{Department of Physics and Astronomy, University of New Mexico, Albuquerque, New Mexico 87106, United States}
\author{Lauren Zundel}
\affiliation{Department of Physics and Astronomy, University of New Mexico, Albuquerque, New Mexico 87106, United States}
\author{Wilton J. M. Kort-Kamp}
\affiliation{Theoretical Division, Los Alamos National Laboratory, Los Alamos, New Mexico 87545, United States}
\author{Diego A. R. Dalvit}
\affiliation{Theoretical Division, Los Alamos National Laboratory, Los Alamos, New Mexico 87545, United States}
\author{Alejandro Manjavacas}
\email[Corresponding author: ]{a.manjavacas@csic.es}
\affiliation{Instituto de \'Optica (IO-CSIC), Consejo Superior de Investigaciones Cient\'ificas, 28006 Madrid, Spain}
\affiliation{Department of Physics and Astronomy, University of New Mexico, Albuquerque, New Mexico 87106, United States}

\date{\today}

\begin{abstract}

The near-field electromagnetic interaction between nanoscale objects produces enhanced radiative heat transfer that can greatly surpass the limits established by far-field black-body radiation.  Here, we present a theoretical framework to describe the temporal dynamics of the radiative heat transfer in ensembles of nanostructures, which is based on the use of an eigenmode expansion of the equations that govern this process. Using this formalism, we identify the fundamental principles that determine the thermalization of collections of nanostructures, revealing general but often unintuitive dynamics. Our results provide an elegant and precise approach to efficiently analyze the temporal dynamics of the near-field radiative heat transfer in systems containing a large number of nanoparticles. 
\end{abstract}

\maketitle


The thermal radiation exchanged between macroscopic bodies separated by macroscopic distances is accurately described by Planck's law \cite{R1965}. However, this description breaks down when the distance between objects or their size becomes significantly smaller than the so-called thermal wavelength, which, for a temperature $T$, is $\lambda_{ T}=2\pi\hbar c/(k_{\rm B} T)$.  In this limit,  the contribution of near-field components of the electromagnetic field \cite{NSH09,RSJ09,OQW11,SGZ14,CHB15,KSF15,STF16,SZF16,SSC19}, together with the strong responses provided by the electromagnetic resonances of nanostructures \cite{DVJ05,VP07,BJD08,NC08,DK10,ama18,ama29,RSM17}, results in enhanced radiative heat transfer (RHT), which can surpass the black-body limit by several orders of magnitude \cite{BMF16,ama49,FTZ18,CG18,BMV20}. 

Near-field RHT is usually described within the framework of fluctuational electrodynamics \cite{PV1971,BMV20}. In particular, when considering collections of nanostructures, a dipole approximation, where each nanoparticle is modeled as a fluctuating dipole, can be exploited \cite{BBJ11,ama18,N14,N15_2,BMV20}. By doing so, it is possible to calculate the power transferred between the different constituents for a particular fixed distribution of temperatures \cite{ama18,BBF15,DZL17,DZL17_1}. However, if one is interested in understanding the temporal evolution of the particle temperatures, this approach presents several disadvantages. Specifically, since the power transferred between the particles depends on their temperatures, which change over time, it is necessary to perform a new calculation at each step in the temporal evolution \cite{MTB13,WW16,SLL20,ama70}. As a result, this approach provides little insight into the fundamental principles that determine the thermalization dynamics, requires separate calculations for each initial condition, and, in addition, can be computationally unfeasible when the number of particles is sufficiently large.


In this letter, we present a different approach to describe the thermalization dynamics of ensembles of nanoparticles. Our approach is based on linearizing the equations that govern the power transferred between the nanoparticles, which allows us to convert them into an eigenvalue problem. By doing so, we find a set of RHT eigenmodes for the ensemble, which completely describe the evolution of the system under any possible initial temperature distribution.  Eigenmode expansions have been applied to a vast range of topics as a way to reveal physical insight \cite{H09_3,paper300,LR17,KRK19,ama66}. Here, using this approach, we identify the general principles that control the thermalization process mediated by near-field RHT, which often give rise to unintuitive behaviors. This insight leads us to explore exotic scenarios, including dynamics in which the temperature of a particle oscillates around the equilibrium temperature as it thermalizes.  The simplicity of this formalism makes it an elegant and efficient method to describe the dynamics of the near-field RHT in ensembles with many nanoparticles.

We consider an ensemble of $N$ nanospheres with radii $R_i$ and temperatures $T_i$, placed at positions $\mathbf{r}_i$ and surrounded by vacuum at $T_0$, which we fix to $300\,$K for the remainder of this letter. We assume that, for all particles, $R_i\ll\lambda_{T}$ and  all interparticle distances $d_{ij}=|\mathbf{r}_i-\mathbf{r}_j|\geq 4\max(R_i,R_j)$, but significantly smaller than $\lambda_{T}$. Therefore, we model the nanoparticles as fluctuating dipoles with electric polarizabilities ${\bm \alpha}_i$. Following previous works \cite{BBJ11,MTB13,N14,N15_2}, the power absorbed by particle $i$ is (see Appendix for details)
\begin{equation}
\mathcal{P}_i =  \sum_{j=1}^{N}  \int_{0}^{\infty}\!\! d\omega f_{ij}(\omega) \left[ n(\omega,T_j)-  n(\omega,T_0)\right], \label{Pi} 
\end{equation}
where $n(\omega,T) = \left[\exp(\hbar \omega / k_{\rm B} T) -1 \right]^{-1}$ is the Bose-Einstein distribution and  $f_{ij}(\omega) = (2\hbar\omega/ \pi){\rm Tr}\left[{\rm Im}\left\{ \mathbf{A}_{ij} {\rm Im}\{{\bm \chi}_j\} \mathbf{C}^{+}_{ij} \right\}\right]$. In this expression, ``+'' represents the conjugate transpose, the trace is taken over Cartesian components, and the different matrices, with dimensions $3N\times3N$, are defined as: $\mathbf{A}=\left[\bm{\mathcal{I}}-\bm{\alpha} \mathbf{G}\right]^{-1}$, $\mathbf{C}=\left(\mathbf{G}+\mathbf{G}^0\right)\mathbf{A}$, and $\bm{\chi}=\bm{\alpha}-\mathbf{G^0}\bm{\alpha}^+\bm{\alpha}$, with $\bm{\mathcal{I}}$ being the identity matrix, ${\bm \alpha}$ a matrix with the polarizabilities, $\mathbf{G}$ the dipole-dipole interaction tensor, and $\mathbf{G}^0=\frac{2i\omega^3 }{3c^3}\bm{\mathcal{I}}$. This model can be generalized to particles with magnetic response by including a magnetic polarizability \cite{ama18,DZL17_1}.

\begin{figure}
\begin{center}
\includegraphics[width=70mm,angle=0]{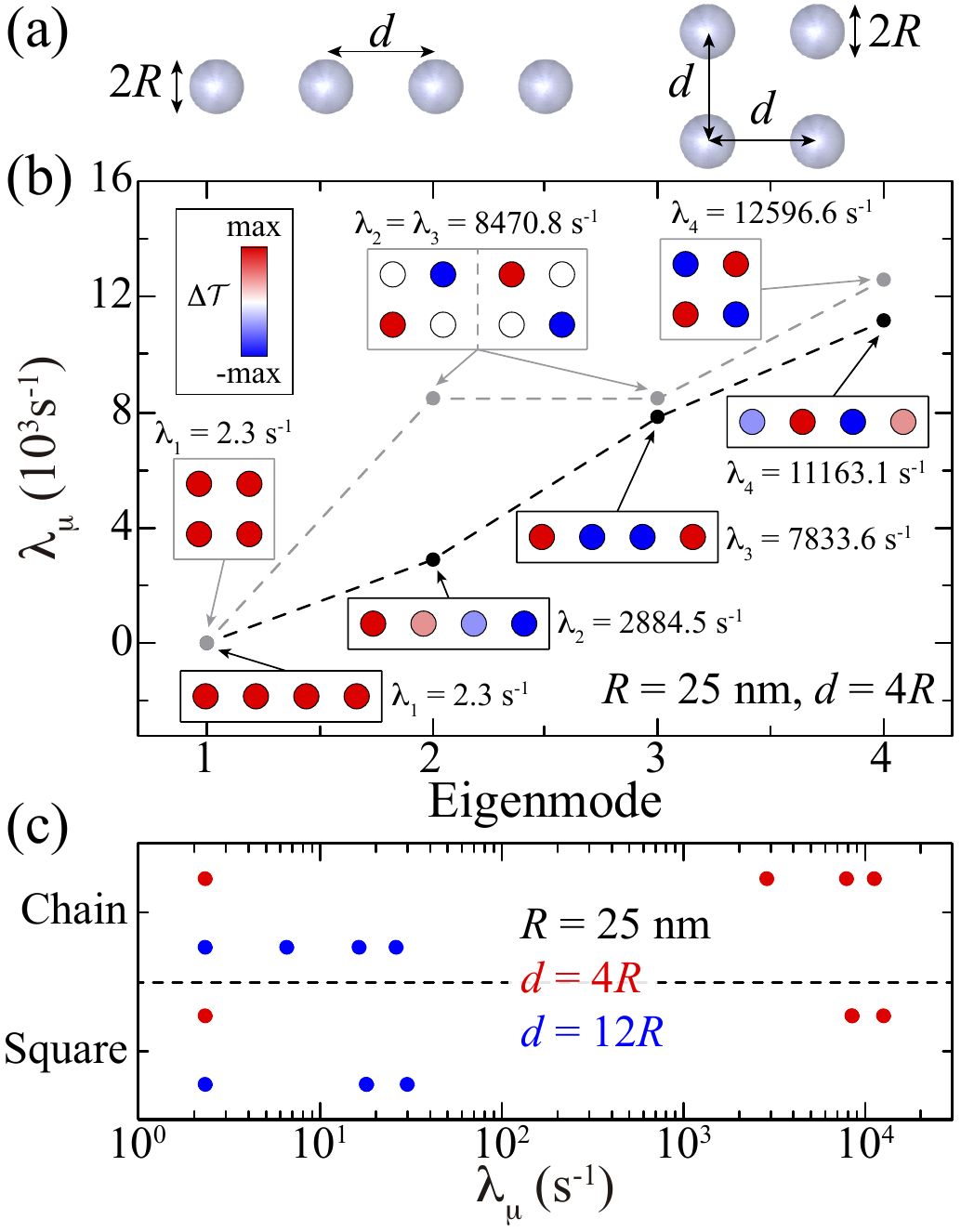}
\caption{(a) Schematics of the systems under study. (b) Decay rates of the RHT eigenmodes of the two systems assuming $R=25\,$nm and $d=4R$. The insets display the components of the RHT eigenmodes and the  value of the associated decay rate. (c) Decay rates for different values of $d$.} \label{fig1}
\end{center}
\end{figure}

The temporal evolution of the temperatures of the nanoparticles is determined by the ratio between the power they absorb $\mathcal{P}_i $ and their heat capacities $\gamma_i$. By expanding $n(\omega,T_j)$ around $T_0$, as $n(\omega, T_j) \approx n(\omega, T_0) + \Delta T_j \partial n(\omega,T) / \partial T|_{T=T_0}$, with $\Delta T_j = T_j-T_0$, we can linearize Eq.~(\ref{Pi}) to obtain the differential equation governing the  evolution of the nanoparticle temperatures,
\begin{equation}
\frac{d }{d t} \Delta \mathbf{T}(t) = - \mathbf{H} \Delta \mathbf{T}(t). \label{Td}
\end{equation}
Here, $\mathbf{H}={\bm \Gamma}^{-1}\mathbf{F}$ is the product of the inverse of a diagonal matrix ${\bm \Gamma}$ containing the heat capacities of the nanoparticles $\gamma_i$ and a symmetric matrix $\mathbf{F}$ with components
\begin{equation}
F_{ij} = - \int_{0}^{\infty} \!\! d\omega f_{ij}(\omega)\left.\frac{\partial n(\omega,T)}{\partial T}\right|_{T=T_0}. \nonumber
\end{equation}
As shown in the Appendix, the structure of $\mathbf{H}$ ensures its diagonalizability. This allows us to write the solution of Eq.~(\ref{Td}) using its eigenvalues $\lambda_{\mu}$ and eigenvectors $\Delta \bm{\mathcal{T}}^{(\mu)}$ as
\begin{equation}
\Delta \mathbf{T} (t) = \sum_{\mu=1}^N a_{\mu} e^{- \lambda_{\mu} t} \Delta \bm{\mathcal{T}}^{(\mu)}, \label{sol}
\end{equation}
where the coefficients $a_{\mu}$ are obtained from the weighted inner product between $\Delta \bm{\mathcal{T}}^{(\mu)}$ and the vector containing the initial temperatures $\Delta \mathbf{T}(0)$ as $a_{\mu}=\sum_{i=1}^N \gamma_i \Delta T_i(0)\Delta \mathcal{T}^{(\mu)}_{i}$, with the eigenvectors satisfying $\sum_{i=1}^N {\gamma_i} \Delta \mathcal{T}^{(\mu)}_i \Delta \mathcal{T}^{(\nu)}_i = \delta_{\mu \nu}$.
Therefore, we conclude from Eq.~(\ref{sol}) that the dynamics of the near-field RHT of an ensemble of nanoparticles can be completely understood by analyzing its RHT eigenmodes and decay rates given, respectively, by the eigenvectors and eigenvalues of $\mathbf{H}$.  Importantly, $\mathbf{H}$ is positive definite (\textit{i.e.}, $\lambda_{\mu}>0$), which ensures that the ensemble thermalizes as $t\rightarrow\infty$.

This approach assumes that the temperature dependence of the material properties of the nanoparticles can be neglected. Furthermore, as discussed in the Appendix, its accuracy improves as $\max(|\Delta T_j|/T_0)$ and $\hbar\omega_0/(k_{\rm B}T_0)$ decrease.  Here, $\omega_0$ represents the characteristic frequency of the electromagnetic response of the nanoparticles. For the systems under consideration, the results of the eigenmode approach have very good agreement with the non-linearized full calculation up to $\max(|\Delta T_j|/T_0)\approx 1/3$, as shown in Fig.~\ref{figS1}.

To illustrate the developed framework, we consider a simple example, although the conclusions we draw are general to any ensemble of nanoparticles. 
In particular, we analyze the two systems depicted in  Fig.~\ref{fig1}(a), consisting of $N=4$ identical SiC spherical nanoparticles arranged in either a chain or a square (see Fig.~\ref{figS2} for a similar analysis of a system with $N=2197$). We obtain the polarizability of the particles from the dipolar Mie coefficient \cite{paper112} using the dielectric function $\varepsilon(\omega)=\varepsilon_{\infty}\left[1+(\omega^2_{\rm L}-\omega^2_{\rm T})/(\omega^2_{\rm T}-\omega^2-i\omega\tau^{-1})\right]$, with $\varepsilon_{\infty}=6.7$, $\hbar\omega_{\rm T}=98.3\,$meV,  $\hbar\omega_{\rm L}=120\,$meV, and $\hbar\tau^{-1}=0.59\,$meV \cite{P1985}. 
Figure~\ref{fig1}(b) analyzes the RHT eigenmodes of the chain (black) and the square (gray) assuming that the particles have a radius $R=25\,$nm and are separated by $d=4R$. 
The chain has four distinct eigenmodes, while the larger symmetry of the square results in two of its modes being degenerate. 
Since particles with the same temperature do not exchange heat with one another, every ensemble, including the two analyzed here, must always have an eigenmode with equal amplitude in all particles. This eigenmode, which we label as $\mu=1$, represents a net transfer of heat between the ensemble and the environment and, as explained below, always has the slowest decay rate. The orthogonality of the eigenmodes forces the rest of them to satisfy $\sum_{i=1}^N \gamma_i \Delta\mathcal{T}^{(\mu>1)}_{i} =0 $, which physically means that they represent processes in which the heat stored in the ensemble remains constant.  Therefore, every eigenmode with $\mu>1$ describes a near-field RHT process among the nanoparticles of the ensemble. 
Examining the components of these eigenmodes, we observe that, as $\mu$ increases, the length scale over which the sign of the components alternates, and hence the near-field RHT occurs, decreases. This is consistent with the increase of the associated decay rate, whose value is dominated by terms proportional to $ (R_iR_j)^3/d_{ij}^6$. In contrast, $\lambda_1$ describes the net radiation exchange between the ensemble and the environment, which scales as $ (R_i/\lambda_{T})^3$. Therefore, for near-field RHT (\textit{i.e.}, $d_{ij}\ll\lambda_T$), $\lambda_1$ always has the smallest value among all of the decay rates, although, as shown in Fig.~\ref{fig1}(c), the difference between $\lambda_1$ and the rest of the decay rates is reduced by increasing the distance between the particles.

\begin{figure}
\begin{center}
\includegraphics[width=70mm,angle=0]{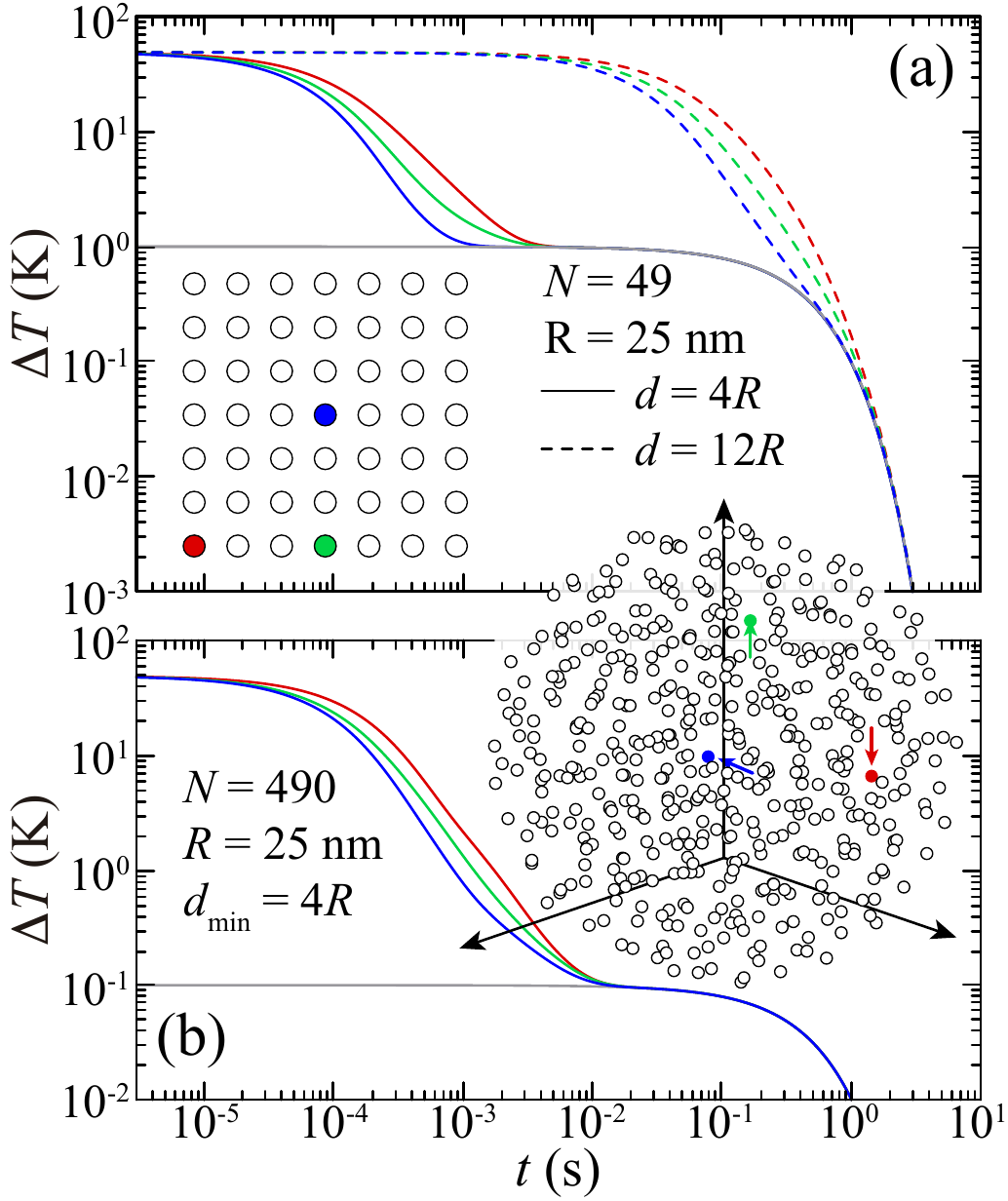}
\caption{ (a) Thermalization dynamics for an array of $N=49$ SiC nanoparticles with $R=25\,$nm under different initial conditions. The colored curves display the evolution of the temperature of the particle of that color in the inset schematics, when such particle is initially at $\Delta T = 49\,$K and the rest at $\Delta T=0\,$K. The gray curves represent the case where all of the particles are initially at $\Delta T=49/N\,$K. In all cases, solid and dashed curves correspond to $d=4R$ and $d=12R$. (b) Same as (a), but for an ensemble of $N=490$ SiC nanoparticles with $R=25\,$nm, randomly distributed inside a spherical volume of radius $600\,$nm with a minimum interparticle distance $d_{\rm min}=4R$ (see schematics). } \label{fig2}
\end{center}
\end{figure}

We know from Eq.~(\ref{sol}) that the thermalization of an ensemble of particles is initially dominated by the eigenmodes with largest decay rates. However, for sufficiently long time, this process is controlled by the first eigenmode, which, as discussed above, has equal amplitude in all particles and, consequently, its decay rate is the smallest. Therefore, in the limit $t\rightarrow\infty$, the thermalization dynamics of a given ensemble  depends exclusively on $a_1 \propto \sum_{i=1}^N \gamma_i \Delta T_i (0)$, or, in other words, the total heat initially stored in it. This gives rise to interesting behaviors, as illustrated in Fig~\ref{fig2}(a). There, we analyze the thermalization dynamics of a square array of $N=49$ identical SiC particles with $R=25\,$nm and $d=4R$ (solid curves). We consider different initial temperature distributions, all of them corresponding to the same value of $a_1$. Specifically, the gray curve displays the evolution of the temperature of the nanoparticles when all of them begin at  $\Delta T = 1\,$K. On the other hand, the colored curves represent different scenarios where only one particle, indicated in the schematics using the same color, is initially hot at $\Delta T=49\,$K. One might anticipate that when all of the particles begin at $\Delta T=1\,$K, the system would thermalize most quickly to the environment. However, as seen in Fig~\ref{fig2}(a), this is not the case. Instead, in all of the scenarios under consideration, all of the particles approach the equilibrium identically as $e^{-\lambda_1 t}$. 

Interestingly, for the scenarios in which only one particle is initially hot, the thermalization process happens over two steps: first, all of the particles converge to $\Delta T=1\,$K and, second, the whole array thermalizes to the environment. This behavior is the result of the large difference between $\lambda_1$ and the rest of the decay rates, as shown in Fig.~\ref{figS3}. Therefore, if such difference is decreased by, for instance, increasing the interparticle distance to $d=12R$, the two-step behavior fades away, as shown by the dashed curves.

Although, so far, we have only considered ordered distributions of particles, our conclusions apply to any arbitrary ensemble of particles. For example, in Fig~\ref{fig2}(b), we consider an ensemble of $N=490$ identical SiC nanoparticles with $R=25\,$nm randomly arranged within a spherical volume of radius $600\,$nm, as shown in the inset. As in Fig.~\ref{fig2}(a), we compare the thermalization process for four different initial conditions; in three of them, one particle, marked in the schematics with the same color as its corresponding curve, begins at $\Delta T = 49\,$K, while, in the fourth (gray curve), all of the particles begin at $\Delta T = 0.1\,$K. As expected, since $a_1$ takes the same value for all of the cases, they all approach the thermalization to the environment identically, despite their very different initial temperature distributions.

\begin{figure}
\begin{center}
\includegraphics[width=70mm,angle=0]{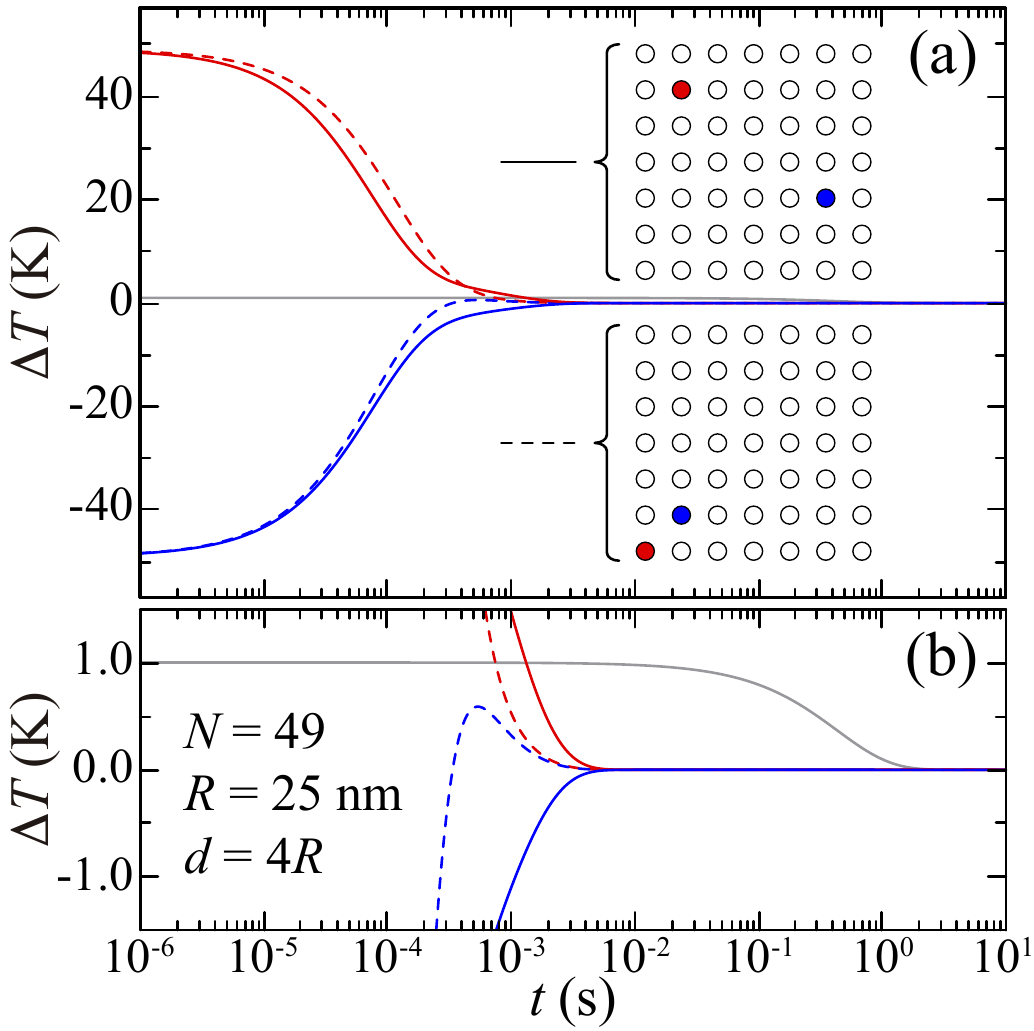}
\caption{ (a) Thermalization dynamics for a hot (red curves) and a cold (blue curves) particle in an array of $N=49$ SiC nanoparticles with $R=25\,$nm and $d=4R$. The red and blue particles are initially at $\Delta T = 49\,$K and $\Delta T = -49\,$K, respectively, while the rest are at $\Delta T=0\,$K. We consider the two cases depicted in the schematics, which are displayed with solid and dashed curves, respectively. For comparison, the gray curve represents the thermalization dynamics when all of the particles are initially at $\Delta T=49/N\,$K. (b) Zoom of (a) around $\Delta T = 0\,$K. } \label{fig3}
\end{center}
\end{figure}

Another interesting scenario to consider is when the initial distribution of temperatures is orthogonal to the first RHT eigenmode and hence $a_1=0$. Physically, this means that, although the system is not thermalized, the total amount of heat initially stored in it is zero. In this case, the thermalization process is governed entirely by the eigenmodes describing the near-field RHT between the particles, since a net transfer of heat to the environment (described by the first RHT eigenmode) is forbidden. To illustrate this, in Fig.~\ref{fig3}(a), we study the thermalization dynamics of the array of Fig.~\ref{fig2}(a) with $d=4R$, for the initial temperature distributions depicted in the insets of Fig.~\ref{fig3}(a). In both of them, one particle begins at $\Delta T = 49\,$K and another at $\Delta T = -49\,$K, while the rest of the array is at $\Delta T=0 \,$K, so $a_1=0$. The corresponding results are displayed using solid and dashed curves, as indicated by the legend, with red and blue colors describing, respectively, the temperature of the hot and cold particles. As expected, in both cases, the thermalization of the array occurs on a time scale $\sim \lambda_2^{-1} \approx 10^{-3}\,$s. This is much faster than the thermalization when all of the nanoparticles begin at $\Delta T = 1\,$K (gray curve),  even though, in that case, the particles have to undergo a temperature change of only $1\,$K [see Fig.~\ref{fig3}(b) for a zoom around $\Delta T=0\,$K]. The reason is, again, the large difference between $\lambda_{\mu>1}$ and $\lambda_1$.

Interestingly, the closer look provided in Fig.~\ref{fig3}(b) reveals an unintuitive behavior:  when the hot and cold particles are next to each other (dashed curves),  the temperature of the initially cold particle rises beyond $\Delta T=0\,$K and subsequently approaches it from above. We attribute this behavior to the difference in the local environment of the two nanoparticles; while the hot one lies on the corner of the array, the cold one is situated in the interior and is therefore surrounded by more particles. This creates an imbalance in the cooling and heating rates of the two particles.

\begin{figure}
\begin{center}
\includegraphics[width=70mm,angle=0]{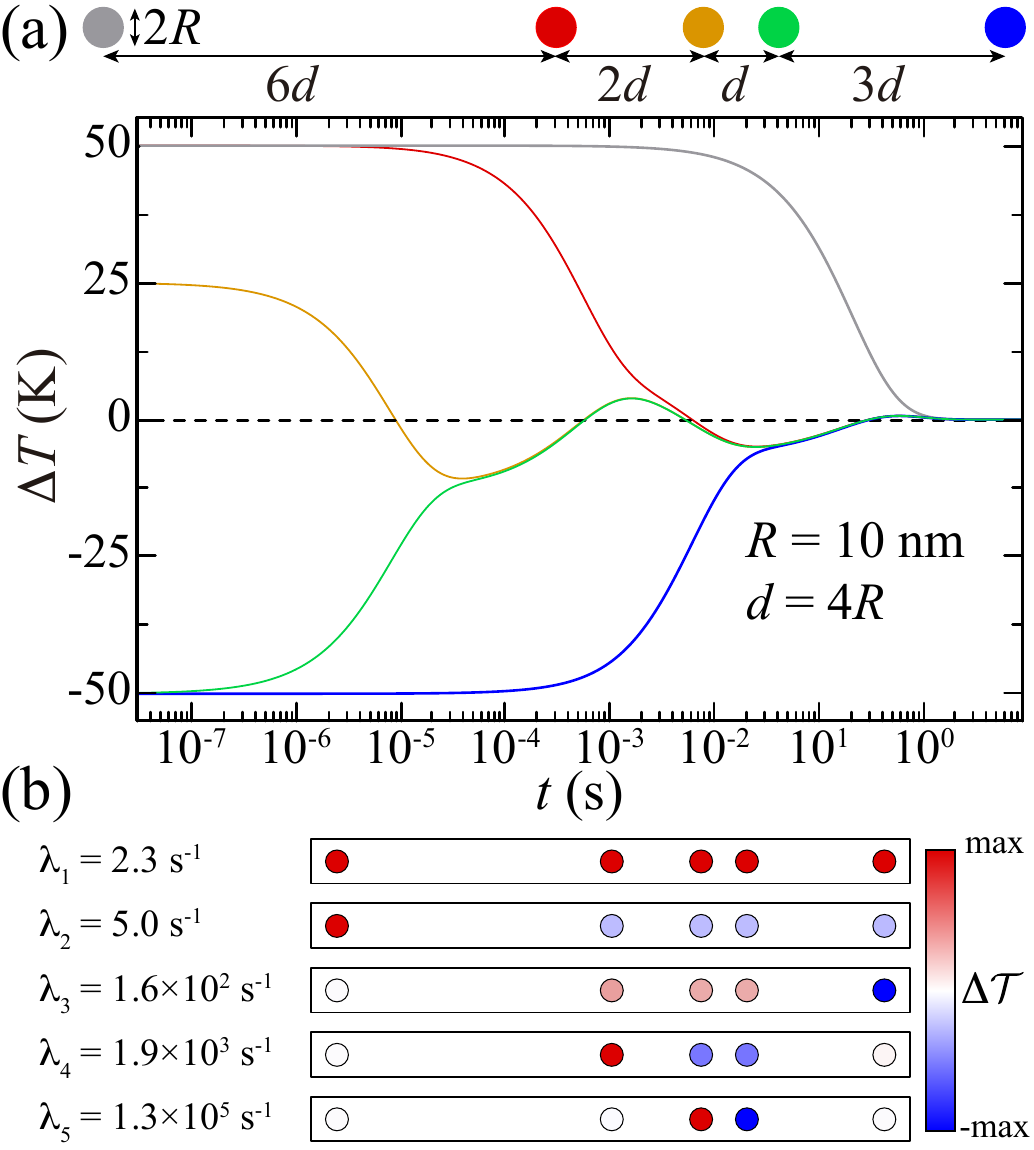}
\caption{(a) Thermalization dynamics for a chain with $N=5$ SiC nanoparticles arranged as shown in the schematics. We assume that $R=10\,$nm, $d=4R$, and the nanoparticles are initially at: $\Delta T = 50$, $50$, $25$, $-50$, and $-50\,$K. (b) RHT eigenmodes of the chain and their associated decay rates.} \label{fig4}
\end{center}
\end{figure}

We can use the RHT eigenmode framework to gain more insight into this oscillatory behavior. To that end, we analyze a simpler system that exhibits similar oscillatory dynamics but in a more pronounced way. In particular, we consider the chain of $N=5$ SiC nanoparticles with $R=10\,$nm and $d=4R$, shown in the schematics of Fig.~\ref{fig4}. The particles are initially at $\Delta T = 50$, $50$, $25$, $-50$, and $-50\,$K. The different curves in Fig.~\ref{fig4}(a) show the evolution of the temperature of the particle with matching color.  As the particles thermalize, their temperatures oscillate around $\Delta T=0\,$K, with the center one (yellow) crossing this value four times throughout the process. The origin of this exotic behavior becomes clear by considering the RHT eigenmodes of the system, which are shown, with their corresponding decay rates, in Fig.~\ref{fig4}(b). Specifically, the initial stage of the thermalization is dominated by the eigenmode with the largest decay rate, which corresponds to a near-field RHT process happening almost exclusively between the center nanoparticle and its nearest neighbor. After that, the contribution of the next fastest eigenmode drives the thermalization of both of those particles with their next-nearest neighbor. This pattern repeats with each successive eigenmode, resulting in the observed oscillatory behavior of $\Delta T$. 

In conclusion, we have presented a theoretical framework to characterize the temporal dynamics of the near-field RHT in arbitrary ensembles of nanoparticles. Our approach is based on an eigenmode expansion of the equations that govern the RHT, obtained upon their linearization. The resulting set of eigenmodes completely characterize the RHT between the constituents of the ensemble and their environment and therefore allow us to express, in a closed form, the evolution of the temperatures of the particles for any initial condition. Exploiting this formalism, we have identified general characteristics of the dynamics of RHT, which often present themselves in unintuitive ways. Specifically, we have shown that an ensemble of nanoparticles beginning with a fixed amount of stored heat always approaches thermalization identically, regardless of how that heat is initially distributed. Similarly, when the total initial heat stored in an ensemble is zero, the system reaches thermal equilibrium faster than the case where there is any initially stored heat. We have also predicted and explained an exotic  behavior in which the temperature of nanoparticles oscillates around the equilibrium value as they thermalize. Our results provide an insightful and computationally efficient approach to study the thermalization dynamics mediated by the near-field RHT, which will facilitate the systematic investigation of the impact that novel phenomena, such as topology \cite{OB20} and nonreciprocity \cite{ZS16,OMB19}, have on this process. Furthermore, this framework can be exploited to analyze the combined transfer of energy and momentum mediated by the fluctuations of the electromagnetic field \cite{ama66}.

\begin{acknowledgments}
This work has been sponsored by the U.S. National Science Foundation (Grant No. DMR-1941680) and the Ministerio de Ciencia, Innovaci\'on y Universidades of Spain (Grant TEM-FLU PID2019-109502GA-I00). L.Z. acknowledges support from the Department of Energy Computational Science Graduate Fellowship (Grant No. DE-SC0020347). D.D. and W.K.K. acknowledge financial support from the Laboratory Directed Research and Development program of Los Alamos National Laboratory under project LDRD 20210327ER.
\end{acknowledgments}

\onecolumngrid
\appendix
\section{Appendix}\label{ap}
\renewcommand{\thefigure}{S\arabic{figure}}
\setcounter{figure}{0} 

\subsection{Derivation of Equation~(\ref{Pi})}

Here, we follow the approach from Refs.~\cite{BBJ11,MTB13,N14,N15_2} to derive Eq.~(\ref{Pi}). Within the dipolar approximation, the power absorbed by particle $i$ in an ensemble with $N$ elements can be written as
\begin{equation}
\mathcal{P}_i=\left\langle\mathbf{E}_i(t)\cdot\frac{\partial\mathbf{p}_i(t)}{\partial t}\right\rangle,\nonumber
\end{equation}
where $\mathbf{E}_i$ is the electric field at the position of the particle, $\mathbf{p}_i$ is its dipole moment, and $\langle\rangle$ stands for the average over thermal fluctuations. Shifting to the frequency domain through the Fourier transform defined as $\mathbf{p}_i(t) = \int_{-\infty}^\infty \!\frac{d\omega}{2\pi}  \mathbf{p}_i(\omega)e^{-i\omega t}$ for the dipole moment, and similarly for the field, $\mathcal{P}_i$ can be rewritten as
\begin{equation}
\mathcal{P}_i=-\int_{-\infty}^{\infty}\frac{d\omega d\omega'}{(2\pi)^2}e^{-i(\omega-\omega')t}i\omega\langle \mathbf{E}_{i}^\ast(\omega')\cdot\mathbf{p}_i(\omega)\rangle,\label{pow}
\end{equation}
where $^*$ represents the complex conjugate. The electric field and the dipole moment appearing in this expression are the self-consistent solutions of the many-body scattering problem for the ensemble with sources $\mathbf{p}\fl$ and $\mathbf{E}\fl$. These sources are the fluctuating dipole and fields arising from the finite temperature of the particles and their environment. By solving this scattering problem, we can write $\mathbf{E}_i$ and $\mathbf{p}_i$ as
\begin{equation}
\mathbf{p}_i=\sum_{j=1}^N\left[\mathbf{A}_{ij}\mathbf{p}_j\fl+\mathbf{B}_{ij}\mathbf{E}_j\fl \right], \qquad \mathbf{E}_i=\sum_{j=1}^N\left[\mathbf{C}_{ij}\mathbf{p}_j\fl+\mathbf{D}_{ij}\mathbf{E}_j\fl\right],\label{EP}
\end{equation}
in terms of the following matrices with dimensions $3N\times 3N$: $\mathbf{A}=\left[\bm{\mathcal{I}}-\bm{\alpha} \mathbf{G}\right]^{-1}$, $\mathbf{B}=\mathbf{A}\bm{\alpha}$, $\mathbf{C}=\left(\mathbf{G}+\mathbf{G}^0\right)\mathbf{A}$, and $\mathbf{D}=\bm{\mathcal{I}}+\mathbf{C}\bm{\alpha}$. Here, $\bm{\mathcal{I}}$ represents the identity matrix, $\bm{\alpha}$ is a matrix that contains the polarizabilities of the nanoparticles, $\mathbf{G}^0=\frac{2i}{3}k^3\bm{\mathcal{I}}$, and $\mathbf{G}$  is the dipole-dipole interaction tensor. The components of $\mathbf{G}$ are zero for $i=j$ and
\begin{equation}
\mathbf{G}_{ij}=\frac{e^{ikd_{ij}}}{d_{ij}^3}\left[(kd_{ij})^2+ikd_{ij}-1\right]\bm{\mathcal{I}}_{3\times3}-\frac{e^{ikd_{ij}}}{d_{ij}^3}\left[(kd_{ij})^2+3ikd_{ij}-3\right]\frac{\mathbf{d}_{ij} \mathbf{d}^{+}_{ij}}{d_{ij}^2},\nonumber
\end{equation}
for $i\neq j$, where $\mathbf{d}_{ij}=\mathbf{r}_i-\mathbf{r}_j$ is the vector describing the distance between particles $i$ and $j$, $\bm{\mathcal{I}}_{3\times3}$ is the $3\times 3$ identity matrix, $k=\omega/c$, and ``+'' represents the conjugate transpose.

Substituting the solutions given in Eq.~(\ref{EP}) into the expression of the power absorbed by dipole $i$ shown in Eq.~(\ref{pow}), we obtain
\begin{equation}
\mathcal{P}_i=-\int_{-\infty}^{\infty}\frac{d\omega d\omega'}{(2\pi)^2}e^{-i(\omega-\omega')t}i\omega \sum_{j,k=1}^N \left\langle\left[\mathbf{C}_{ij}\mathbf{p}_j\fl+\mathbf{D}_{ij}\mathbf{E}_j\fl\right]^+\left[\mathbf{A}_{ik}\mathbf{p}_k\fl+\mathbf{B}_{ik}\mathbf{E}_k\fl\right]\right\rangle. \nonumber
\end{equation}
In order to perform the average over fluctuations, we use the fluctuation-dissipation theorem \cite{R1959_2,ama7} (FDT), which takes the form 
\begin{equation}
\langle \mathbf{p}^{\rm fl}_{i}(\omega) \mathbf{p}_{j}^{{\rm fl} +}(\omega')\rangle=4\pi\hbar\delta(\omega-\omega'){\rm Im}\{\bm{\chi}_i\}\delta_{ij}\left[n(\omega,T_i)+\frac{1}{2}\right],\nonumber
\end{equation}
for the dipole fluctuations and
\begin{equation}
\langle \mathbf{E}^{\rm fl}_{i}(\omega)\mathbf{E}_{j}^{{\rm fl}+}(\omega')\rangle=4\pi\hbar\delta(\omega-\omega'){\rm Im}\{\mathbf{G}_{ij}+\mathbf{G}^0_{ij}\}\left[n(\omega,T_0)+\frac{1}{2}\right],\nonumber
\end{equation}
for the electric field fluctuations. In these expressions, $n(\omega,T_i)$ represents the Bose-Einstein distribution for temperature $T_i$, with $T_0$ being the temperature of the environment, and $\bm{\chi}=\bm{\alpha}-\mathbf{G^0}\bm{\alpha}^+\bm{\alpha}$. Then, using these expressions and noting that any cross terms involving dipole and field fluctuations vanish, since they are uncorrelated, we obtain
\begin{equation}
\mathcal{P}_i=\int_{0}^{\infty}d\omega \sum_{j=1}^N \left[ f_{ij} n(\omega,T_j) + f^0_{ij}n(\omega,T_0)\right],\nonumber
\end{equation}
where  $f_{ij}= (2\hbar\omega/ \pi){\rm Tr}\left[{\rm Im}\{\mathbf{A}_{ij}{\rm Im}\{\bm{\chi}_j\}\mathbf{C}_{ij}^+\}\right]$ and $f^0_{ij}=(2\hbar\omega/ \pi){\rm Tr}\left[ \sum_{j'=1}^N {\rm Im}\{\mathbf{B}_{ij}{\rm Im}\{\mathbf{G}_{jj'}+\mathbf{G}^0_{jj'}\}\mathbf{D}_{ij'}^+\}\right]$, with the trace taken over Cartesian components. 
Finally, since the power absorbed by particle $i$ must vanish when the temperatures of all particles are equal to $T_0$, regardless of the actual value of $T_0$, we have that $\sum_{j=1}^Nf^0_{ij}=-\sum_{j=1}^Nf_{ij}$, which yields Eq.~(\ref{Pi}).

\subsection{Diagonalizability of the matrix $\mathbf{H}$}

As explained in the main text, the thermalization dynamics of an ensemble of nanoparticles, induced by the near-field radiative heat transfer (RHT), can be characterized by analyzing the matrix $\mathbf{H}$. This matrix is the product of a positive definite diagonal matrix ${\bm \Gamma}^{-1}$, whose entries are the inverse of the heat capacities $\gamma_i$ of the different nanoparticles, and a symmetric matrix $\mathbf{F}$. When all of the particles in the ensemble are identical, the matrix $\mathbf{H}$ is real and symmetric and therefore diagonalizable by the spectral theorem \cite{FIS03}. The situation is more complicated when the particles in the array have different heat capacities. In this case, despite still being real, $\mathbf{H}$ is not symmetric because ${\bm \Gamma}^{-1}$ does not commute with $\mathbf{F}$. However, $\mathbf{H}$ is still diagonalizable, as we show in the following. First, it is worth noting that $\sqrt{{\bm \Gamma}}$ exists and is symmetric because ${\bm \Gamma}$ is a diagonal positive definite matrix. Then, let us consider the similarity transformation
\begin{equation}
\sqrt{{\bm \Gamma}}  \mathbf{H} \sqrt{{\bm \Gamma}} \, ^{-1} = \mathbf{H}_{\rm S}
,  \label{sim_transform}
\end{equation}
where we have introduced the matrix $\mathbf{H}_{\rm S} = \sqrt{{\bm \Gamma}}\, ^{-1} \mathbf{F} \sqrt{{\bm \Gamma}}\, ^{-1}$. Clearly, $\mathbf{H}_{\rm S}$ is symmetric because it is equal to its transpose. Therefore, $\mathbf{H}_{\rm S}$ is diagonalizable and, consequently, has a complete set of eigenvalues $\lambda_\mu $ and corresponding eigenvectors $ \Delta\bm{\tau}^{(\mu)}$ that satisfy 
\begin{equation}
\mathbf{H}_{\rm s} \Delta \bm{\tau}^{(\mu)} = \lambda_\mu \Delta \bm{\tau}^{(\mu)}. \nonumber
\end{equation}
Furthermore, from Eq.~(\ref{sim_transform}), it is clear that $\mathbf{H}$ has the same eigenvalues as $\mathbf{H}_{\rm S}$ and eigenvectors given by $\Delta\bm{\mathcal{T}}^{(\mu)} = \sqrt{{\bm \Gamma}}^{-1} \Delta\bm{\tau}^{(\mu)} $. Although, in general, these vectors are not orthogonal under the usual inner product, they are orthogonal using an inner product weighted by the heat capacities
\begin{equation}
\sum_{i=1}^N {\gamma_i} \Delta \mathcal{T}^{(\mu)}_i \Delta \mathcal{T}^{(\nu)}_i = \sum_{i=1}^N  \Delta \tau^{(\mu)}_i \Delta \tau^{(\nu)}_i = \delta_{\mu\nu}, \nonumber
\end{equation}
where we have used the orthonormality of $\Delta\bm{\tau}^{(\mu)} $.
Then, the solution of Eq.~(2) of the main paper is given by 
\begin{equation}
\Delta \mathbf{T}(t) = \sum_{\mu=1}^N a_{\mu} e^{- \lambda_{\mu} t} \Delta \bm{\mathcal{T}}^{(\mu)}, \nonumber
\end{equation}
where the coefficients $a_{\mu}$ are defined as
\begin{equation} 
a_{\mu}=\sum_{i=1}^N \gamma_i \Delta T_i (0) \Delta \mathcal{T}_i^{(\mu)}. \nonumber
\end{equation}

\begin{figure}
\begin{center}
\includegraphics[width=130mm,angle=0]{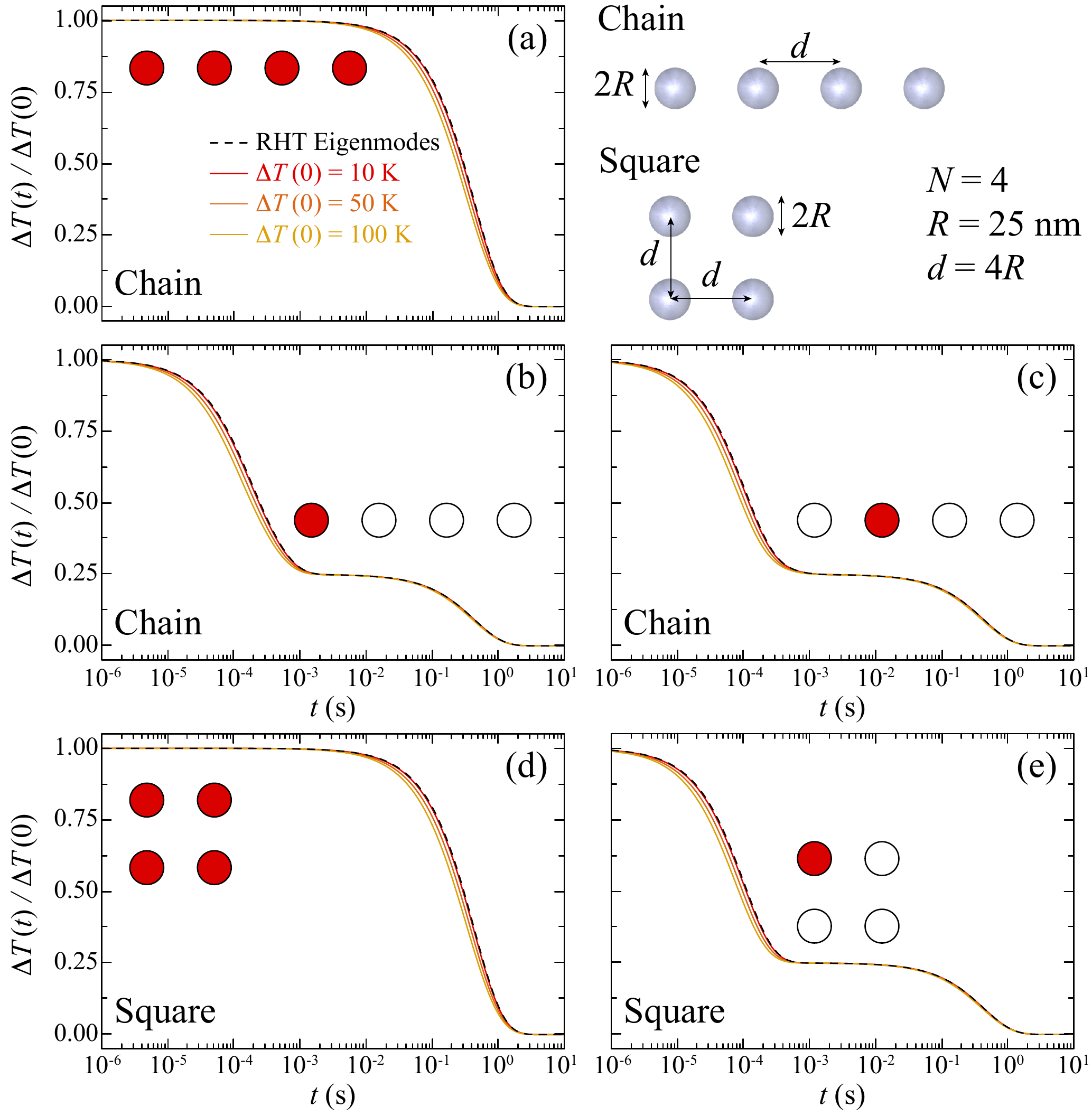}
\caption{ (a-c) Comparison of the thermalization dynamics obtained using the near-field RHT eigenmode formalism (dashed curves) with equivalent results obtained using the non-linearized full calculation (solid curves) for a chain of $N=4$ identical SiC nanoparticles with $R=25\,$nm and $d=4R$, as shown in the schematics. The different colored curves correspond to different values of $\Delta T(0)$ for the particles, as indicated by the legend, while the different panels correspond to different initial temperature distributions. Specifically, in panel (a) all of the particles begin at the same $\Delta T(0)$, while in panels (b) and (c) only one of the particles (see inset) is initially hot and the rest are thermalized to the environment. In each case, we plot the evolution of the temperature of the particle(s) that is (are) initially hot. (d-e) Same as (a-c), but for a square arrangement of the particles. In all cases, the agreement between the RHT eigenmode formalism and the  non-linearized full calculation is excellent. As stated in the main paper, the accuracy of the RHT eigenmode formalism improves as $\max(|\Delta T_j|/T_0)$ and $\hbar\omega_0/(k_{\rm B}T_0)$ decrease. This can be understood by analyzing the relative error in the expansion of the factor involving the Bose-Einstein distributions $n(\omega,T_j)-n(\omega,T_0)$, which can be approximated by 
\begin{minipage}{\linewidth}
\begin{equation*}
\epsilon \approx \frac{\Delta T_j}{2} \frac{n^{\prime\prime}(\omega_0,T_0)}{n^{\prime}(\omega_0,T_0)},
\end{equation*}
\end{minipage}\vspace{1em}
where the prime denotes a derivative with respect to $T$. Notice that both derivatives are evaluated at $T_0$ and $\omega_0$. The latter is the characteristic frequency of the electromagnetic response of the nanoparticles, which is expected to dominate the integral over frequencies that leads to $F_{ij}$. Explicitly calculating the first and second derivatives of the Bose-Einstein distribution, we have
\begin{minipage}{\linewidth}
\begin{equation*}
\epsilon \approx \frac{1}{2}\frac{\Delta T_j}{T_0} \left( \frac{\hbar\omega_0}{k_{\rm B}T_0} \coth\left[\frac{\hbar\omega_0}{2k_{\rm B}T_0}\right] - 2\right).
\end{equation*}
\end{minipage}\vspace{1em}
Therefore, it is clear that the relative error decreases as $\Delta T_j/T_0$ and $\hbar\omega_0/(k_{\rm B}T_0)$ are reduced.
} \label{figS1}
\end{center}
\end{figure}

\begin{figure}
\begin{center}
\includegraphics[width=130mm,angle=0]{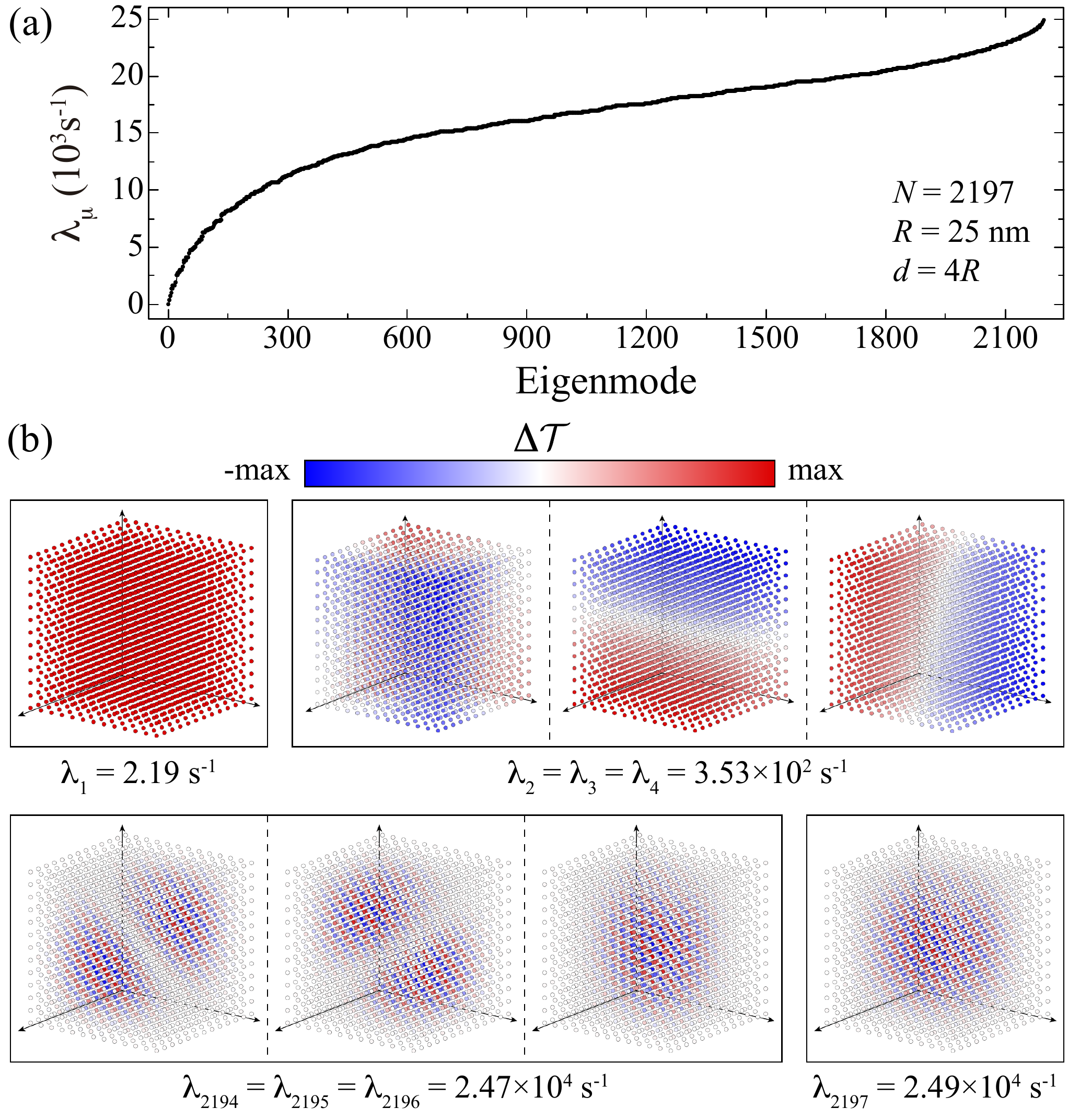}
\caption{(a) Decay rates of the RHT eigenmodes of a cubic array of $N=13^3=2197$ identical SiC nanoparticles with $R=25\,$nm and $d=4R$. (b) First and last four RHT eigenmodes of the system and their corresponding decay rates. Eigenmodes $2-4$ and $2194-2196$ each have a threefold degeneracy due to the symmetry of the ensemble. As discussed in main paper, the first eigenmode, whose components are all equal, represents the net transfer of heat between the ensemble and the environment. On the other hand, the last eigenmode represents the fastest near-field RHT process and therefore its components alternate sign over neighboring particles, with the amplitude decaying away from the center of the ensemble.} \label{figS2}
\end{center}
\end{figure}

\begin{figure}
\begin{center}
\includegraphics[width=130mm,angle=0]{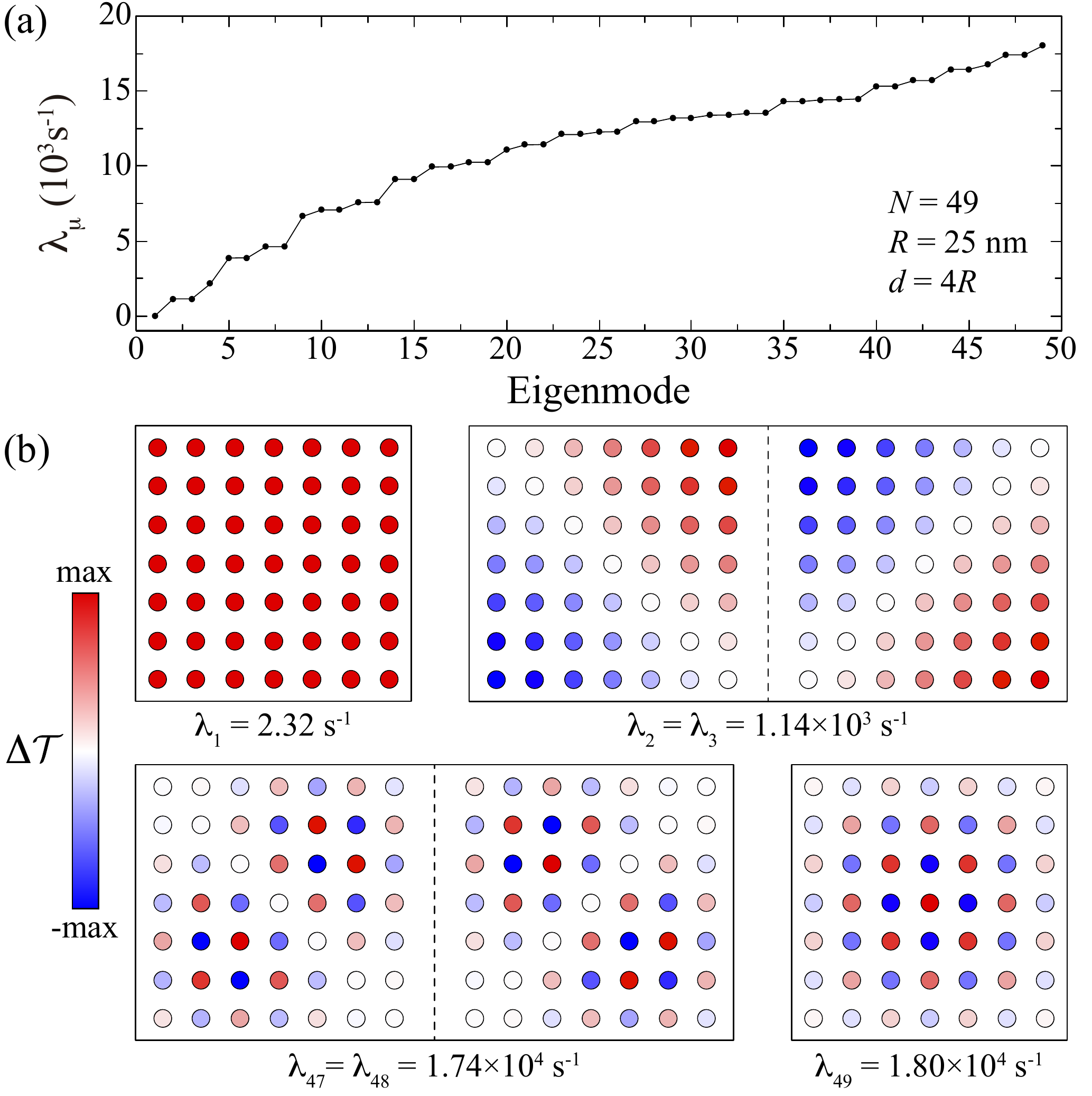}
\caption{(a) Decay rates of the RHT eigenmodes of the system considered in Figures~2 and 3 of the main paper, consisting of a square array of $N=49$ identical SiC nanoparticles with $R=25\,$nm and $d=4R$. (b)  First and last three RHT eigenmodes of the system and their corresponding decay rates. Eigenmodes $2$ and $3$, as well as $47$ and $48$, have a twofold degeneracy due to the symmetry of the ensemble. As in Figure~\ref{figS2}, the first eigenmode has equal amplitude across all particles, while, in the last one, nearest neighbors alternate signs and the amplitude decreases closer to the edges of the array.} \label{figS3}
\end{center}
\end{figure}

\newpage

\bibliographystyle{apsrev}

\begin{thebibliography}{47}
\expandafter\ifx\csname natexlab\endcsname\relax\def\natexlab#1{#1}\fi
\expandafter\ifx\csname bibnamefont\endcsname\relax
  \def\bibnamefont#1{#1}\fi
\expandafter\ifx\csname bibfnamefont\endcsname\relax
  \def\bibfnamefont#1{#1}\fi
\expandafter\ifx\csname citenamefont\endcsname\relax
  \def\citenamefont#1{#1}\fi
\expandafter\ifx\csname url\endcsname\relax
  \def\url#1{\texttt{#1}}\fi
\expandafter\ifx\csname urlprefix\endcsname\relax\def\urlprefix{URL }\fi
\providecommand{\bibinfo}[2]{#2}
\providecommand{\eprint}[2][]{\url{#2}}

\bibitem[{\citenamefont{Reif}(1965)}]{R1965}
\bibinfo{author}{\bibfnamefont{F.}~\bibnamefont{Reif}},
  \emph{\bibinfo{title}{Fundamentals of Statistical and Thermal Physics}}
  (\bibinfo{publisher}{McGraw-Hill}, \bibinfo{address}{New York},
  \bibinfo{year}{1965}).

\bibitem[{\citenamefont{Narayanaswamy et~al.}(2009)\citenamefont{Narayanaswamy,
  Shen, Hu, Chen, and Chen}}]{NSH09}
\bibinfo{author}{\bibfnamefont{A.}~\bibnamefont{Narayanaswamy}},
  \bibinfo{author}{\bibfnamefont{S.}~\bibnamefont{Shen}},
  \bibinfo{author}{\bibfnamefont{L.}~\bibnamefont{Hu}},
  \bibinfo{author}{\bibfnamefont{X.}~\bibnamefont{Chen}}, \bibnamefont{and}
  \bibinfo{author}{\bibfnamefont{G.}~\bibnamefont{Chen}},
  \bibinfo{journal}{Appl.\ Phys.\ A} \textbf{\bibinfo{volume}{96}},
  \bibinfo{pages}{357} (\bibinfo{year}{2009}).

\bibitem[{\citenamefont{Rousseau et~al.}(2009)\citenamefont{Rousseau, Siria,
  Jourdan, Volz, Comin, Chevrier, and Greffet}}]{RSJ09}
\bibinfo{author}{\bibfnamefont{E.}~\bibnamefont{Rousseau}},
  \bibinfo{author}{\bibfnamefont{A.}~\bibnamefont{Siria}},
  \bibinfo{author}{\bibfnamefont{G.}~\bibnamefont{Jourdan}},
  \bibinfo{author}{\bibfnamefont{S.}~\bibnamefont{Volz}},
  \bibinfo{author}{\bibfnamefont{F.}~\bibnamefont{Comin}},
  \bibinfo{author}{\bibfnamefont{J.}~\bibnamefont{Chevrier}}, \bibnamefont{and}
  \bibinfo{author}{\bibfnamefont{J.~J.} \bibnamefont{Greffet}},
  \bibinfo{journal}{Nat.\ Photon.} \textbf{\bibinfo{volume}{3}},
  \bibinfo{pages}{514} (\bibinfo{year}{2009}).

\bibitem[{\citenamefont{Ottens et~al.}(2011)\citenamefont{Ottens, Quetschke,
  Wise, Alemi, Lundock, Mueller, Reitze, Tanner, and Whiting}}]{OQW11}
\bibinfo{author}{\bibfnamefont{R.~S.} \bibnamefont{Ottens}},
  \bibinfo{author}{\bibfnamefont{V.}~\bibnamefont{Quetschke}},
  \bibinfo{author}{\bibfnamefont{S.}~\bibnamefont{Wise}},
  \bibinfo{author}{\bibfnamefont{A.~A.} \bibnamefont{Alemi}},
  \bibinfo{author}{\bibfnamefont{R.}~\bibnamefont{Lundock}},
  \bibinfo{author}{\bibfnamefont{G.}~\bibnamefont{Mueller}},
  \bibinfo{author}{\bibfnamefont{D.~H.} \bibnamefont{Reitze}},
  \bibinfo{author}{\bibfnamefont{D.~B.} \bibnamefont{Tanner}},
  \bibnamefont{and} \bibinfo{author}{\bibfnamefont{B.~F.}
  \bibnamefont{Whiting}}, \bibinfo{journal}{Phys.\ Rev.\ Lett.}
  \textbf{\bibinfo{volume}{107}}, \bibinfo{pages}{014301}
  (\bibinfo{year}{2011}).

\bibitem[{\citenamefont{St-Gelais et~al.}(2014)\citenamefont{St-Gelais, Guha,
  Zhu, Fan, and Lipson}}]{SGZ14}
\bibinfo{author}{\bibfnamefont{R.}~\bibnamefont{St-Gelais}},
  \bibinfo{author}{\bibfnamefont{B.}~\bibnamefont{Guha}},
  \bibinfo{author}{\bibfnamefont{L.}~\bibnamefont{Zhu}},
  \bibinfo{author}{\bibfnamefont{S.}~\bibnamefont{Fan}}, \bibnamefont{and}
  \bibinfo{author}{\bibfnamefont{M.}~\bibnamefont{Lipson}},
  \bibinfo{journal}{Nano\ Lett.} \textbf{\bibinfo{volume}{14}},
  \bibinfo{pages}{6971} (\bibinfo{year}{2014}).

\bibitem[{\citenamefont{Chalabi et~al.}(2015)\citenamefont{Chalabi, Hasman, and
  Brongersma}}]{CHB15}
\bibinfo{author}{\bibfnamefont{H.}~\bibnamefont{Chalabi}},
  \bibinfo{author}{\bibfnamefont{E.}~\bibnamefont{Hasman}}, \bibnamefont{and}
  \bibinfo{author}{\bibfnamefont{M.~L.} \bibnamefont{Brongersma}},
  \bibinfo{journal}{Phys.\ Rev.\ B} \textbf{\bibinfo{volume}{91}},
  \bibinfo{pages}{014302} (\bibinfo{year}{2015}).

\bibitem[{\citenamefont{Kim et~al.}(2015)\citenamefont{Kim, Song,
  Fern\'andez-Hurtado, Lee, Jeong, Cui, Thompson, Feist, Reid, Garc\'{i}a-Vidal
  et~al.}}]{KSF15}
\bibinfo{author}{\bibfnamefont{K.}~\bibnamefont{Kim}},
  \bibinfo{author}{\bibfnamefont{B.}~\bibnamefont{Song}},
  \bibinfo{author}{\bibfnamefont{V.}~\bibnamefont{Fern\'andez-Hurtado}},
  \bibinfo{author}{\bibfnamefont{W.}~\bibnamefont{Lee}},
  \bibinfo{author}{\bibfnamefont{W.}~\bibnamefont{Jeong}},
  \bibinfo{author}{\bibfnamefont{L.}~\bibnamefont{Cui}},
  \bibinfo{author}{\bibfnamefont{D.}~\bibnamefont{Thompson}},
  \bibinfo{author}{\bibfnamefont{J.}~\bibnamefont{Feist}},
  \bibinfo{author}{\bibfnamefont{M.~T.~H.} \bibnamefont{Reid}},
  \bibinfo{author}{\bibfnamefont{F.~J.} \bibnamefont{Garc\'{i}a-Vidal}},
  \bibnamefont{et~al.}, \bibinfo{journal}{Nature}
  \textbf{\bibinfo{volume}{528}}, \bibinfo{pages}{387} (\bibinfo{year}{2015}).

\bibitem[{\citenamefont{Song et~al.}(2016)\citenamefont{Song, Thompson,
  Fiorino, Ganjeh, Reddy, and Meyhofer}}]{STF16}
\bibinfo{author}{\bibfnamefont{B.}~\bibnamefont{Song}},
  \bibinfo{author}{\bibfnamefont{D.}~\bibnamefont{Thompson}},
  \bibinfo{author}{\bibfnamefont{A.}~\bibnamefont{Fiorino}},
  \bibinfo{author}{\bibfnamefont{Y.}~\bibnamefont{Ganjeh}},
  \bibinfo{author}{\bibfnamefont{P.}~\bibnamefont{Reddy}}, \bibnamefont{and}
  \bibinfo{author}{\bibfnamefont{E.}~\bibnamefont{Meyhofer}},
  \bibinfo{journal}{Nat.\ Nanotechnol.} \textbf{\bibinfo{volume}{11}},
  \bibinfo{pages}{509} (\bibinfo{year}{2016}).

\bibitem[{\citenamefont{St-Gelais et~al.}(2016)\citenamefont{St-Gelais, Zhu,
  Fan, and Lipson}}]{SZF16}
\bibinfo{author}{\bibfnamefont{R.}~\bibnamefont{St-Gelais}},
  \bibinfo{author}{\bibfnamefont{L.}~\bibnamefont{Zhu}},
  \bibinfo{author}{\bibfnamefont{S.}~\bibnamefont{Fan}}, \bibnamefont{and}
  \bibinfo{author}{\bibfnamefont{M.}~\bibnamefont{Lipson}},
  \bibinfo{journal}{Nat.\ Nanotechnol.} \textbf{\bibinfo{volume}{11}},
  \bibinfo{pages}{515} (\bibinfo{year}{2016}).

\bibitem[{\citenamefont{Shi et~al.}(2019)\citenamefont{Shi, Sun, Chen, He, Bao,
  Evans, and He}}]{SSC19}
\bibinfo{author}{\bibfnamefont{K.}~\bibnamefont{Shi}},
  \bibinfo{author}{\bibfnamefont{Y.}~\bibnamefont{Sun}},
  \bibinfo{author}{\bibfnamefont{Z.}~\bibnamefont{Chen}},
  \bibinfo{author}{\bibfnamefont{N.}~\bibnamefont{He}},
  \bibinfo{author}{\bibfnamefont{F.}~\bibnamefont{Bao}},
  \bibinfo{author}{\bibfnamefont{J.}~\bibnamefont{Evans}}, \bibnamefont{and}
  \bibinfo{author}{\bibfnamefont{S.}~\bibnamefont{He}}, \bibinfo{journal}{Nano\
  Lett.} \textbf{\bibinfo{volume}{19}}, \bibinfo{pages}{8082}
  (\bibinfo{year}{2019}).

\bibitem[{\citenamefont{Domingues et~al.}(2005)\citenamefont{Domingues, Volz,
  Joulain, and Greffet}}]{DVJ05}
\bibinfo{author}{\bibfnamefont{G.}~\bibnamefont{Domingues}},
  \bibinfo{author}{\bibfnamefont{S.}~\bibnamefont{Volz}},
  \bibinfo{author}{\bibfnamefont{K.}~\bibnamefont{Joulain}}, \bibnamefont{and}
  \bibinfo{author}{\bibfnamefont{J.~J.} \bibnamefont{Greffet}},
  \bibinfo{journal}{Phys.\ Rev.\ Lett.} \textbf{\bibinfo{volume}{94}},
  \bibinfo{pages}{085901} (\bibinfo{year}{2005}).

\bibitem[{\citenamefont{Volokitin and Persson}(2007)}]{VP07}
\bibinfo{author}{\bibfnamefont{A.~I.} \bibnamefont{Volokitin}}
  \bibnamefont{and} \bibinfo{author}{\bibfnamefont{B.~N.~J.}
  \bibnamefont{Persson}}, \bibinfo{journal}{Rev.\ Mod.\ Phys.}
  \textbf{\bibinfo{volume}{79}}, \bibinfo{pages}{1291} (\bibinfo{year}{2007}).

\bibitem[{\citenamefont{Ben-Abdallah et~al.}(2008)\citenamefont{Ben-Abdallah,
  Joulain, Drevillon, and Le~Goff}}]{BJD08}
\bibinfo{author}{\bibfnamefont{P.}~\bibnamefont{Ben-Abdallah}},
  \bibinfo{author}{\bibfnamefont{K.}~\bibnamefont{Joulain}},
  \bibinfo{author}{\bibfnamefont{J.}~\bibnamefont{Drevillon}},
  \bibnamefont{and} \bibinfo{author}{\bibfnamefont{C.}~\bibnamefont{Le~Goff}},
  \bibinfo{journal}{Phys.\ Rev.\ B} \textbf{\bibinfo{volume}{77}},
  \bibinfo{pages}{075417} (\bibinfo{year}{2008}).

\bibitem[{\citenamefont{Narayanaswamy and Chen}(2008)}]{NC08}
\bibinfo{author}{\bibfnamefont{A.}~\bibnamefont{Narayanaswamy}}
  \bibnamefont{and} \bibinfo{author}{\bibfnamefont{G.}~\bibnamefont{Chen}},
  \bibinfo{journal}{Phys.\ Rev.\ B} \textbf{\bibinfo{volume}{77}},
  \bibinfo{pages}{075125} (\bibinfo{year}{2008}).

\bibitem[{\citenamefont{Dedkov and Kyasov}(2010)}]{DK10}
\bibinfo{author}{\bibfnamefont{G.~V.} \bibnamefont{Dedkov}} \bibnamefont{and}
  \bibinfo{author}{\bibfnamefont{A.~A.} \bibnamefont{Kyasov}},
  \bibinfo{journal}{J.\ Comput.\ Theor.\ Nanosci.}
  \textbf{\bibinfo{volume}{7}}, \bibinfo{pages}{2019} (\bibinfo{year}{2010}).

\bibitem[{\citenamefont{Manjavacas and Garc\'{\i}a~de Abajo}(2012)}]{ama18}
\bibinfo{author}{\bibfnamefont{A.}~\bibnamefont{Manjavacas}} \bibnamefont{and}
  \bibinfo{author}{\bibfnamefont{F.~J.} \bibnamefont{Garc\'{\i}a~de Abajo}},
  \bibinfo{journal}{Phys.\ Rev.\ B} \textbf{\bibinfo{volume}{86}},
  \bibinfo{pages}{075466} (\bibinfo{year}{2012}).

\bibitem[{\citenamefont{Manjavacas et~al.}(2014)\citenamefont{Manjavacas,
  Thongrattanasiri, Greffet, and {Garc\'{\i}a de Abajo}}}]{ama29}
\bibinfo{author}{\bibfnamefont{A.}~\bibnamefont{Manjavacas}},
  \bibinfo{author}{\bibfnamefont{S.}~\bibnamefont{Thongrattanasiri}},
  \bibinfo{author}{\bibfnamefont{J.~J.} \bibnamefont{Greffet}},
  \bibnamefont{and} \bibinfo{author}{\bibfnamefont{F.~J.}
  \bibnamefont{{Garc\'{\i}a de Abajo}}}, \bibinfo{journal}{Appl.\ Phys.\ Lett.}
  \textbf{\bibinfo{volume}{105}}, \bibinfo{pages}{211102}
  (\bibinfo{year}{2014}).

\bibitem[{\citenamefont{Ramirez et~al.}(2017)\citenamefont{Ramirez, Shen, and
  McGaughey}}]{RSM17}
\bibinfo{author}{\bibfnamefont{F.~V.} \bibnamefont{Ramirez}},
  \bibinfo{author}{\bibfnamefont{S.}~\bibnamefont{Shen}}, \bibnamefont{and}
  \bibinfo{author}{\bibfnamefont{A.~J.~H.} \bibnamefont{McGaughey}},
  \bibinfo{journal}{Phys.\ Rev.\ B} \textbf{\bibinfo{volume}{96}},
  \bibinfo{pages}{165427} (\bibinfo{year}{2017}).

\bibitem[{\citenamefont{Bernardi et~al.}(2016)\citenamefont{Bernardi, Milovich,
  and Francoeur}}]{BMF16}
\bibinfo{author}{\bibfnamefont{M.~P.} \bibnamefont{Bernardi}},
  \bibinfo{author}{\bibfnamefont{D.}~\bibnamefont{Milovich}}, \bibnamefont{and}
  \bibinfo{author}{\bibfnamefont{M.}~\bibnamefont{Francoeur}},
  \bibinfo{journal}{Nat.\ Commun.} \textbf{\bibinfo{volume}{7}},
  \bibinfo{pages}{12900} (\bibinfo{year}{2016}).

\bibitem[{\citenamefont{Yu et~al.}(2017{\natexlab{a}})\citenamefont{Yu,
  Manjavacas, and {Garc\'{\i}a de Abajo}}}]{ama49}
\bibinfo{author}{\bibfnamefont{R.}~\bibnamefont{Yu}},
  \bibinfo{author}{\bibfnamefont{A.}~\bibnamefont{Manjavacas}},
  \bibnamefont{and} \bibinfo{author}{\bibfnamefont{F.~J.}
  \bibnamefont{{Garc\'{\i}a de Abajo}}}, \bibinfo{journal}{Nat.\ Commun.}
  \textbf{\bibinfo{volume}{8}}, \bibinfo{pages}{2}
  (\bibinfo{year}{2017}{\natexlab{a}}).

\bibitem[{\citenamefont{Fiorino et~al.}(2018)\citenamefont{Fiorino, Thompson,
  Zhu, Mittapally, Biehs, Bezencenet, El-Bondry, Bansropun, Ben-Abdallah,
  Meyhofer et~al.}}]{FTZ18}
\bibinfo{author}{\bibfnamefont{A.}~\bibnamefont{Fiorino}},
  \bibinfo{author}{\bibfnamefont{D.}~\bibnamefont{Thompson}},
  \bibinfo{author}{\bibfnamefont{L.}~\bibnamefont{Zhu}},
  \bibinfo{author}{\bibfnamefont{R.}~\bibnamefont{Mittapally}},
  \bibinfo{author}{\bibfnamefont{S.-A.} \bibnamefont{Biehs}},
  \bibinfo{author}{\bibfnamefont{O.}~\bibnamefont{Bezencenet}},
  \bibinfo{author}{\bibfnamefont{N.}~\bibnamefont{El-Bondry}},
  \bibinfo{author}{\bibfnamefont{S.}~\bibnamefont{Bansropun}},
  \bibinfo{author}{\bibfnamefont{P.}~\bibnamefont{Ben-Abdallah}},
  \bibinfo{author}{\bibfnamefont{E.}~\bibnamefont{Meyhofer}},
  \bibnamefont{et~al.}, \bibinfo{journal}{ACS\ Nano}
  \textbf{\bibinfo{volume}{12}}, \bibinfo{pages}{5774} (\bibinfo{year}{2018}).

\bibitem[{\citenamefont{Cuevas and Garca-Vidal}(2018)}]{CG18}
\bibinfo{author}{\bibfnamefont{J.~C.} \bibnamefont{Cuevas}} \bibnamefont{and}
  \bibinfo{author}{\bibfnamefont{F.~J.} \bibnamefont{Garca-Vidal}},
  \bibinfo{journal}{ACS\ Photonics} \textbf{\bibinfo{volume}{5}},
  \bibinfo{pages}{3896} (\bibinfo{year}{2018}).

\bibitem[{\citenamefont{Biehs et~al.}(2020)\citenamefont{Biehs, Messina,
  Venkataram, Rodriguez, Cuevas, and Ben-Abdallah}}]{BMV20}
\bibinfo{author}{\bibfnamefont{S.-A.} \bibnamefont{Biehs}},
  \bibinfo{author}{\bibfnamefont{R.}~\bibnamefont{Messina}},
  \bibinfo{author}{\bibfnamefont{P.~S.} \bibnamefont{Venkataram}},
  \bibinfo{author}{\bibfnamefont{A.~W.} \bibnamefont{Rodriguez}},
  \bibinfo{author}{\bibfnamefont{J.~C.} \bibnamefont{Cuevas}},
  \bibnamefont{and}
  \bibinfo{author}{\bibfnamefont{P.}~\bibnamefont{Ben-Abdallah}},
  \bibinfo{journal}{0} \textbf{\bibinfo{volume}{0}},
  \bibinfo{pages}{arXiv:2007.05604v1} (\bibinfo{year}{2020}).

\bibitem[{\citenamefont{Polder and {Van Hove}}(1971)}]{PV1971}
\bibinfo{author}{\bibfnamefont{D.}~\bibnamefont{Polder}} \bibnamefont{and}
  \bibinfo{author}{\bibfnamefont{M.}~\bibnamefont{{Van Hove}}},
  \bibinfo{journal}{Phys.\ Rev.\ B} \textbf{\bibinfo{volume}{4}},
  \bibinfo{pages}{3303} (\bibinfo{year}{1971}).

\bibitem[{\citenamefont{Ben-Abdallah et~al.}(2011)\citenamefont{Ben-Abdallah,
  Biehs, and Joulain}}]{BBJ11}
\bibinfo{author}{\bibfnamefont{P.}~\bibnamefont{Ben-Abdallah}},
  \bibinfo{author}{\bibfnamefont{S.~A.} \bibnamefont{Biehs}}, \bibnamefont{and}
  \bibinfo{author}{\bibfnamefont{K.}~\bibnamefont{Joulain}},
  \bibinfo{journal}{Phys.\ Rev.\ Lett.} \textbf{\bibinfo{volume}{107}},
  \bibinfo{pages}{114301} (\bibinfo{year}{2011}).

\bibitem[{\citenamefont{Nikbakht}(2014)}]{N14}
\bibinfo{author}{\bibfnamefont{M.}~\bibnamefont{Nikbakht}},
  \bibinfo{journal}{J.\ Appl.\ Phys.} \textbf{\bibinfo{volume}{116}},
  \bibinfo{pages}{094307} (\bibinfo{year}{2014}).

\bibitem[{\citenamefont{Nikbakht}(2015)}]{N15_2}
\bibinfo{author}{\bibfnamefont{M.}~\bibnamefont{Nikbakht}},
  \bibinfo{journal}{{EPL} (Europhysics Letters)}
  \textbf{\bibinfo{volume}{110}}, \bibinfo{pages}{14004}
  (\bibinfo{year}{2015}).

\bibitem[{\citenamefont{Ben-Abdallah et~al.}(2015)\citenamefont{Ben-Abdallah,
  Belarouci, Frechette, and Biehs}}]{BBF15}
\bibinfo{author}{\bibfnamefont{P.}~\bibnamefont{Ben-Abdallah}},
  \bibinfo{author}{\bibfnamefont{A.}~\bibnamefont{Belarouci}},
  \bibinfo{author}{\bibfnamefont{L.}~\bibnamefont{Frechette}},
  \bibnamefont{and} \bibinfo{author}{\bibfnamefont{S.-A.} \bibnamefont{Biehs}},
  \bibinfo{journal}{App.\ Phys.\ Lett.} \textbf{\bibinfo{volume}{107}},
  \bibinfo{pages}{053109} (\bibinfo{year}{2015}).

\bibitem[{\citenamefont{Dong et~al.}(2017{\natexlab{a}})\citenamefont{Dong,
  Zhao, and Liu}}]{DZL17}
\bibinfo{author}{\bibfnamefont{J.}~\bibnamefont{Dong}},
  \bibinfo{author}{\bibfnamefont{J.}~\bibnamefont{Zhao}}, \bibnamefont{and}
  \bibinfo{author}{\bibfnamefont{L.}~\bibnamefont{Liu}}, \bibinfo{journal}{J.
  Quant.\ Spectrosc.\ Radiat.\ Transfer} \textbf{\bibinfo{volume}{197}},
  \bibinfo{pages}{114 } (\bibinfo{year}{2017}{\natexlab{a}}).

\bibitem[{\citenamefont{Dong et~al.}(2017{\natexlab{b}})\citenamefont{Dong,
  Zhao, and Liu}}]{DZL17_1}
\bibinfo{author}{\bibfnamefont{J.}~\bibnamefont{Dong}},
  \bibinfo{author}{\bibfnamefont{J.}~\bibnamefont{Zhao}}, \bibnamefont{and}
  \bibinfo{author}{\bibfnamefont{L.}~\bibnamefont{Liu}},
  \bibinfo{journal}{Phys.\ Rev.\ B} \textbf{\bibinfo{volume}{95}},
  \bibinfo{pages}{125411} (\bibinfo{year}{2017}{\natexlab{b}}).

\bibitem[{\citenamefont{Messina et~al.}(2013)\citenamefont{Messina, Tschikin,
  Biehs, and Ben-Abdallah}}]{MTB13}
\bibinfo{author}{\bibfnamefont{R.}~\bibnamefont{Messina}},
  \bibinfo{author}{\bibfnamefont{M.}~\bibnamefont{Tschikin}},
  \bibinfo{author}{\bibfnamefont{S.-A.} \bibnamefont{Biehs}}, \bibnamefont{and}
  \bibinfo{author}{\bibfnamefont{P.}~\bibnamefont{Ben-Abdallah}},
  \bibinfo{journal}{Phys.\ Rev.\ B} \textbf{\bibinfo{volume}{88}},
  \bibinfo{pages}{104307} (\bibinfo{year}{2013}).

\bibitem[{\citenamefont{Wang and Wu}(2016)}]{WW16}
\bibinfo{author}{\bibfnamefont{Y.}~\bibnamefont{Wang}} \bibnamefont{and}
  \bibinfo{author}{\bibfnamefont{J.}~\bibnamefont{Wu}}, \bibinfo{journal}{AIP
  Adv.} \textbf{\bibinfo{volume}{6}}, \bibinfo{pages}{025104}
  (\bibinfo{year}{2016}).

\bibitem[{\citenamefont{Song et~al.}(2020)\citenamefont{Song, Lu, Li, Zhang,
  Hu, Zhou, and Cheng}}]{SLL20}
\bibinfo{author}{\bibfnamefont{J.}~\bibnamefont{Song}},
  \bibinfo{author}{\bibfnamefont{L.}~\bibnamefont{Lu}},
  \bibinfo{author}{\bibfnamefont{B.}~\bibnamefont{Li}},
  \bibinfo{author}{\bibfnamefont{B.}~\bibnamefont{Zhang}},
  \bibinfo{author}{\bibfnamefont{R.}~\bibnamefont{Hu}},
  \bibinfo{author}{\bibfnamefont{X.}~\bibnamefont{Zhou}}, \bibnamefont{and}
  \bibinfo{author}{\bibfnamefont{Q.}~\bibnamefont{Cheng}},
  \bibinfo{journal}{Int.\ J.\ Heat Mass Transf.}
  \textbf{\bibinfo{volume}{150}}, \bibinfo{pages}{119346}
  (\bibinfo{year}{2020}).

\bibitem[{\citenamefont{Zundel and Manjavacas}(2020)}]{ama70}
\bibinfo{author}{\bibfnamefont{L.}~\bibnamefont{Zundel}} \bibnamefont{and}
  \bibinfo{author}{\bibfnamefont{A.}~\bibnamefont{Manjavacas}},
  \bibinfo{journal}{Phys.\ Rev.\ Applied} \textbf{\bibinfo{volume}{13}},
  \bibinfo{pages}{054054} (\bibinfo{year}{2020}).

\bibitem[{\citenamefont{Hussein}(2009)}]{H09_3}
\bibinfo{author}{\bibfnamefont{M.~I.} \bibnamefont{Hussein}},
  \bibinfo{journal}{Proc.\ R.\ Soc.\ A} \textbf{\bibinfo{volume}{465}},
  \bibinfo{pages}{2825} (\bibinfo{year}{2009}).

\bibitem[{\citenamefont{Yu et~al.}(2017{\natexlab{b}})\citenamefont{Yu,
  Liz-Marz\'an, and {Garc\'ia de Abajo}}}]{paper300}
\bibinfo{author}{\bibfnamefont{R.}~\bibnamefont{Yu}},
  \bibinfo{author}{\bibfnamefont{L.~M.} \bibnamefont{Liz-Marz\'an}},
  \bibnamefont{and} \bibinfo{author}{\bibfnamefont{F.~J.}
  \bibnamefont{{Garc\'ia de Abajo}}}, \bibinfo{journal}{Chem.\ Soc.\ Rev.}
  \textbf{\bibinfo{volume}{46}}, \bibinfo{pages}{6710}
  (\bibinfo{year}{2017}{\natexlab{b}}).

\bibitem[{\citenamefont{Lu and Raz}(2017)}]{LR17}
\bibinfo{author}{\bibfnamefont{Z.}~\bibnamefont{Lu}} \bibnamefont{and}
  \bibinfo{author}{\bibfnamefont{O.}~\bibnamefont{Raz}},
  \bibinfo{journal}{Proc.\ Natl.\ Acad.\ Sci.} \textbf{\bibinfo{volume}{114}},
  \bibinfo{pages}{5083} (\bibinfo{year}{2017}).

\bibitem[{\citenamefont{Kocharovsky et~al.}(2019)\citenamefont{Kocharovsky,
  Reynolds, and Kocharovsky}}]{KRK19}
\bibinfo{author}{\bibfnamefont{V.~V.} \bibnamefont{Kocharovsky}},
  \bibinfo{author}{\bibfnamefont{C.~B.} \bibnamefont{Reynolds}},
  \bibnamefont{and} \bibinfo{author}{\bibfnamefont{V.~V.}
  \bibnamefont{Kocharovsky}}, \bibinfo{journal}{Phys.\ Rev.\ A}
  \textbf{\bibinfo{volume}{100}}, \bibinfo{pages}{053854}
  (\bibinfo{year}{2019}).

\bibitem[{\citenamefont{Sanders et~al.}(2019)\citenamefont{Sanders, Kort-Kamp,
  Dalvit, and Manjavacas}}]{ama66}
\bibinfo{author}{\bibfnamefont{S.}~\bibnamefont{Sanders}},
  \bibinfo{author}{\bibfnamefont{W.~J.~M.} \bibnamefont{Kort-Kamp}},
  \bibinfo{author}{\bibfnamefont{D.~A.~R.} \bibnamefont{Dalvit}},
  \bibnamefont{and}
  \bibinfo{author}{\bibfnamefont{A.}~\bibnamefont{Manjavacas}},
  \bibinfo{journal}{Commun.\ Phys.} \textbf{\bibinfo{volume}{2}},
  \bibinfo{pages}{71} (\bibinfo{year}{2019}).

\bibitem[{\citenamefont{Myroshnychenko
  et~al.}(2008)\citenamefont{Myroshnychenko, {Rodr\'{\i}guez-Fern\'andez},
  Pastoriza-Santos, Funston, Novo, Mulvaney, {Liz-Marz\'an}, and {Garc\'{\i}a
  de Abajo}}}]{paper112}
\bibinfo{author}{\bibfnamefont{V.}~\bibnamefont{Myroshnychenko}},
  \bibinfo{author}{\bibfnamefont{J.}~\bibnamefont{{Rodr\'{\i}guez-Fern\'andez}}},
  \bibinfo{author}{\bibfnamefont{I.}~\bibnamefont{Pastoriza-Santos}},
  \bibinfo{author}{\bibfnamefont{A.~M.} \bibnamefont{Funston}},
  \bibinfo{author}{\bibfnamefont{C.}~\bibnamefont{Novo}},
  \bibinfo{author}{\bibfnamefont{P.}~\bibnamefont{Mulvaney}},
  \bibinfo{author}{\bibfnamefont{L.~M.} \bibnamefont{{Liz-Marz\'an}}},
  \bibnamefont{and} \bibinfo{author}{\bibfnamefont{F.~J.}
  \bibnamefont{{Garc\'{\i}a de Abajo}}}, \bibinfo{journal}{Chem.\ Soc.\ Rev.}
  \textbf{\bibinfo{volume}{37}}, \bibinfo{pages}{1792} (\bibinfo{year}{2008}).

\bibitem[{\citenamefont{Palik}(1985)}]{P1985}
\bibinfo{author}{\bibfnamefont{E.~D.} \bibnamefont{Palik}},
  \emph{\bibinfo{title}{Handbook of Optical Constants of Solids}}
  (\bibinfo{publisher}{Academic Press}, \bibinfo{address}{San Diego},
  \bibinfo{year}{1985}).

\bibitem[{\citenamefont{Ott and Biehs}(2020)}]{OB20}
\bibinfo{author}{\bibfnamefont{A.}~\bibnamefont{Ott}} \bibnamefont{and}
  \bibinfo{author}{\bibfnamefont{S.-A.} \bibnamefont{Biehs}},
  \bibinfo{journal}{Phys.\ Rev.\ B} \textbf{\bibinfo{volume}{102}},
  \bibinfo{pages}{115417} (\bibinfo{year}{2020}).

\bibitem[{\citenamefont{Zhu and Fan}(2016)}]{ZS16}
\bibinfo{author}{\bibfnamefont{L.}~\bibnamefont{Zhu}} \bibnamefont{and}
  \bibinfo{author}{\bibfnamefont{S.}~\bibnamefont{Fan}},
  \bibinfo{journal}{Phys.\ Rev.\ Lett.} \textbf{\bibinfo{volume}{117}},
  \bibinfo{pages}{134303} (\bibinfo{year}{2016}).

\bibitem[{\citenamefont{Ott et~al.}(2019)\citenamefont{Ott, Messina,
  Ben-Abdallah, and Biehs}}]{OMB19}
\bibinfo{author}{\bibfnamefont{A.}~\bibnamefont{Ott}},
  \bibinfo{author}{\bibfnamefont{R.}~\bibnamefont{Messina}},
  \bibinfo{author}{\bibfnamefont{P.}~\bibnamefont{Ben-Abdallah}},
  \bibnamefont{and} \bibinfo{author}{\bibfnamefont{S.-A.} \bibnamefont{Biehs}},
  \bibinfo{journal}{Appl.\ Phys.\ Lett.} \textbf{\bibinfo{volume}{114}},
  \bibinfo{pages}{163105} (\bibinfo{year}{2019}).

\bibitem[{\citenamefont{Rytov}(1959)}]{R1959_2}
\bibinfo{author}{\bibfnamefont{S.~M.} \bibnamefont{Rytov}},
  \emph{\bibinfo{title}{Theory of Electric Fluctuations and Thermal Radiation}}
  (\bibinfo{publisher}{Air Force Cambridge Research Center},
  \bibinfo{address}{Bedford, MA}, \bibinfo{year}{1959}).

\bibitem[{\citenamefont{Manjavacas and {Garc\'{\i}a de Abajo}}(2010)}]{ama7}
\bibinfo{author}{\bibfnamefont{A.}~\bibnamefont{Manjavacas}} \bibnamefont{and}
  \bibinfo{author}{\bibfnamefont{F.~J.} \bibnamefont{{Garc\'{\i}a de Abajo}}},
  \bibinfo{journal}{Phys.\ Rev.\ Lett.} \textbf{\bibinfo{volume}{105}},
  \bibinfo{pages}{113601} (\bibinfo{year}{2010}).

\bibitem[{\citenamefont{Friedberg et~al.}(2003)\citenamefont{Friedberg, Insel,
  and Spence}}]{FIS03}
\bibinfo{author}{\bibfnamefont{S.}~\bibnamefont{Friedberg}},
  \bibinfo{author}{\bibfnamefont{A.}~\bibnamefont{Insel}}, \bibnamefont{and}
  \bibinfo{author}{\bibfnamefont{L.}~\bibnamefont{Spence}},
  \emph{\bibinfo{title}{Linear Algebra}}, Featured Titles for Linear Algebra
  (Advanced) Series (\bibinfo{publisher}{Pearson Education},
  \bibinfo{year}{2003}).

\end{thebibliography}

\end{document}